\documentclass[twocolumn,showpacs,preprintnumbers,amsmath,amssymb]{revtex4}

\usepackage{graphicx}
\usepackage[tight]{subfigure}
\usepackage{dcolumn}
\usepackage{bm}

\begin{document}

  \title{Electronic properties of H on vicinal Pt surfaces: A first-principles study}

  \author{T. Vehvil\"ainen}
  \author{P. Salo}
    \email{Petri.Salo@tkk.fi}
  \author{T. Ala-Nissila}
  \altaffiliation[Also at ]{Department of Physics, Brown University, Providence,
     Rhode Island 02912-1843
  }%
  \affiliation{%
    Department of Applied Physics, Helsinki University of Technology, P.O.\ Box
    1100, FI-02015 TKK, Espoo, Finland
  }%

  \author{S.C. Ying}
  \affiliation{%
     Department of Physics, Brown University, Providence,
     Rhode Island 02912-1843
  }%

\date{\today}

\begin{abstract}
In this work, we use the first-principle density-functional approach
to study the electronic structure of a H atom adsorbed on the ideal
Pt(111) and vicinal Pt(211) and Pt(331) surfaces. Full
three-dimensional potential-energy surfaces (3D PES's) as well as
local electronic density of states on various adsorption sites are
obtained. The results show that the steps modify the PES
considerably. The effect is nonlocal and extends into the region of
the (111) terraces. We also find that different type of steps have
different kind of influence on the PES when compared to the one of
the ideal Pt(111) surface. The full 3D PES's calculated in this work
provide a starting point for the theoretical study of vibrational
and diffusive dynamics of H adatoms adsorbed on these vicinal
surfaces.
\end{abstract}



\maketitle


\section{Introduction}

Systems in which hydrogen adatoms interact with transition metal
surfaces have been intensively studied both experimentally and
theoretically \cite{Nis05} since they have many important
applications. For example, hydrogen adsorbed on platinum surfaces
or on platinum nano-particles constitutes an important application
in catalysis of heterogeneous hydrogenation reactions, in the
combustion of hydrocarbons, and in proton exchange membrane fuel
cells.

On a more fundamental level, we need to understand basic dynamical
properties of the adsorbed hydrogen such as vibrational excitations
and diffusion coefficients in order to understand the chemical
reactivity and catalytic activity on a more microscopic level. A
particularly intriguing characteristic of H adatoms on metal
surfaces is that they may exhibit distinct quantum-mechanical
behavior at relatively high temperatures as compared to heavier
elements or molecular adsorbates. In particular, H adatoms on metal
surfaces can diffuse either through classical activated hopping or
quantum-mechanical tunneling processes. At low enough temperatures
many features of hydrogen diffusion on metal surfaces can only be
explained by quantum effects, which are due to the small mass and
delocalization of elemental H
\cite{Nis05,Chr79,Pus83,Pus85,Dan95,Cao97,Lau00,Nob01,Bad01,Sun04,Lei06,Rom06}.

The best studied system in this respect is H on Pt surfaces, as
there have been a number of experimental and theoretical studies on
the vibrational and diffusive dynamics of this system
\cite{Chr76a,Chr76b,Bar79,Ric87,DiW92,Hor99,Gra99,Bad02,Bad03,Nob01,
Rom06,Fei87,Fei97,Ols99,Pij00,Nob00,Pap00,Wat01,Kal01,Leg04,For05,Fag05,Hon05,Fea06,
Gee00,Ols04,Ols04a,Zhe04,Lup05,McC05,Lud06,Zhe06,Ols08,Gro08}.
Recently, the complete three-dimensional potential-energy surface
(3D PES) for H/Pt(111) has  been calculated from first-principles
density-functional theory (DFT) approach. The vibrational
excitations for this system are now well understood in terms of
quantum transitions within a vibrational band structure
\cite{Bad02,Bad03}. The experimental data for diffusion of H on
Pt(111) are still controversial, as different techniques yield
different results \cite{Gra99,Zhe04,Zhe06}. This can be qualitative
understood from the fact that the different techniques study the
diffusion of H adatoms at different length scales. Thus, the
influence of disorder plays a significantly different role in
different experiments. This is particularly important in the
low-temperature regime, where the quantum nature of H adatom motion
is important. Here the role of disorder can lead to the localization
of the wave-function of the H adatom that suppresses diffusion on a
large length scale, such as in the optical diffraction grating
experiments \cite{Cao97,Zhe04,Zhe06}. Again, for quantitative
studies of diffusion, a necessary starting input is the microscopic
PES and the ensuing vibrational band structures.

Realistic surfaces are hardly ever ideal and may contain defects
such as isolated vacancies, dislocation lines or steps. On such
surfaces, the defect sites are often catalytically very active.
Stepped metal surfaces with a regular step spacing can be easily
prepared by cutting a single crystal in a direction vicinal to a
high-symmetry plane.  Hydrogen on stepped platinum surfaces has been
studied
\cite{Gee00,Ols04,Ols04a,Zhe04,Lup05,McC05,Lud06,Zhe06,Ols08,Gro08}
as the presence of step edges increases the dissociation and
sticking of hydrogen \cite{Chr76b}. For instance, Gee {\it et al}.\
\cite{Gee00} have suggested that step sites are responsible for the
low-energy dissociative adsorption of H$_2$ and the high-energy
dissociation of H$_2$ is associated with the (111) terrace sites on
the Pt(533) surface. McCormack {\it et al}.\ \cite{McC05} carried
out density-functional calculations to study the dissociative
adsorption of H$_2$ on the Pt(211) surface, and their results are in
good agreement with the experiment results for H$_2$ on Pt(533)
\cite{Gee00}. Recently, Olsen {\it et al}.\ \cite{Ols08} have
studied H$_2$ dissociation on the Pt(211) stepped surface using
quantum-dynamics calculations. Their results show enhancement of
dissociation at low collision energies in qualitative accordance
with recent experiments \cite{Gro08}.

Zheng {\it et al}.\ \cite{Zhe04,Zhe06} have studied diffusion
dynamics of H on stepped Pt(111) surfaces experimentally using a
linear optical diffraction technique. They found faster diffusion
perpendicular to the steps than that on a flat Pt(111) surface
\cite{Zhe04}. This cannot be explained within the lattice gas model
indicating that the effect of steps have to be nonlocal in the sense
that it extends to the terrace sites. Motivated by this result, in
this work we study the electronic properties of a hydrogen atom, in
particular the PES of H adsorbed on the vicinal Pt(211) and Pt(331)
surfaces using DFT calculations.  We calculate the full 3D PES for
the two vicinal surfaces and present a comprehensive comparison
between the two types of steps on the H/Pt(111) system. The results
show that the steps indeed modify the PES considerably even in the
region of (111) terraces and the different type of steps have
different kind of effect on the PES when compared to the PES of the
flat Pt(111) surface. The PES's obtained can be used in the future
to study the effect of steps {\it e.g.}\ on the vibrational band
structure, and classical and quantum diffusion of H on vicinal Pt
surfaces.

This paper is organized as follows. In Section II we describe the
computational details, in Section III we present the results and
discuss the influence of steps on the behavior of H on Pt
surfaces, and finally in Section IV we give some concluding
remarks based on the results obtained.


\section{Methods}

We have done the DFT calculations using the Vienna {\it Ab-initio}
Simulation Package (VASP) \cite{vasp} with the generalized-gradient
approximation (GGA) of PW91 \cite{pw91} and with projector-augmented
wave (PAW) potentials \cite{PAW}. The cutoff energy for the
plane-wave expansion is 450 eV. For the irreducible Brillouin-zone
integration we use Monkhorst-Pack \cite{Mon76} 13$\times$13$\times$1
\textbf{k}-point sampling  for Pt(111) and 6$\times$6$\times$1 mesh
for Pt(211) and Pt(331), together with a smearing of 0.2 eV for the
Methfessel-Paxton occupation function which significantly shortens
the computing time for transition metal compared to other existing
methods \cite{Met89}. The energy convergence with respect to the
\textbf{k}-points and cutoff energy has been tested and based on
these tests the estimated accuracy of the energies is of the order
of 0.01 eV and the accuracy of structures is within 0.01 {\AA}. On
the other hand, the error due to the choice of the GGA etc.\ is
about 0.02 eV as will be shown in the case of H on Pt(111) later on.

The crystal structure of Pt is face-centered cubic (fcc) and the
calculated lattice constant of Pt is 3.987 {\AA}, the experimental
value being 3.924 {\AA} \cite{Ash76}. We use a supercell which
consists of 4, 12 and 11 atomic layers, corresponding 4, 36 and 33
atoms in the supercell for the Pt(111), Pt(211), and Pt(331)
systems, respectively. Thus, the surface unit cell consists of
1$\times$1 atom for the Pt(111) surface and 3$\times$3 atoms for the
Pt(211) and Pt(331) surfaces. The distance between the slabs is at
least 12 {\AA} which is sufficient to eliminate the interaction
between them.

Two hydrogen atoms form a H$_2$ molecule whose calculated bond
length is 0.75 {\AA} and binding energy $E_{bin}=E_{{\rm
H}_2}-2E_{\rm H}=-4.55$ eV. These values are in an excellent
agreement with experimental values \cite{webelements}. Based on the
data for the energy \emph{vs.}\ bond length of a hydrogen molecule
we can say that H atoms do not interact with each other if the
distance between them is over 7 {\AA} (see \cite{webpage}). When we
construct our Pt(211) and Pt(331) supercells such that the H-H
distance is larger than this value we study effectively a single H
atom interacting with the surface. This is very important since the
H-H interaction can significantly influence the energy of the saddle
point \cite{Zhe04}. In our calculations, the distance between the H
atoms on the same (111) terrace, but in different supercells, is
8.46 {\AA}.

In the slab relaxation the bottom-most layer is fixed in the Pt(111)
calculations and in the Pt(211) and Pt(331) calculations six bottom
layers are fixed to the bulk coordinates.  After the relaxation all
the Pt atom positions are fixed and no Pt atom movements are allowed
during the calculations for the hydrogen atom on a surface. The
relaxations of Pt atoms due to H are small and can therefore be
neglected \cite{Bad02}.

\begin{figure}
\hspace*{1mm}(a) \hspace*{38mm} (b)\\
\includegraphics[width=4.0cm]{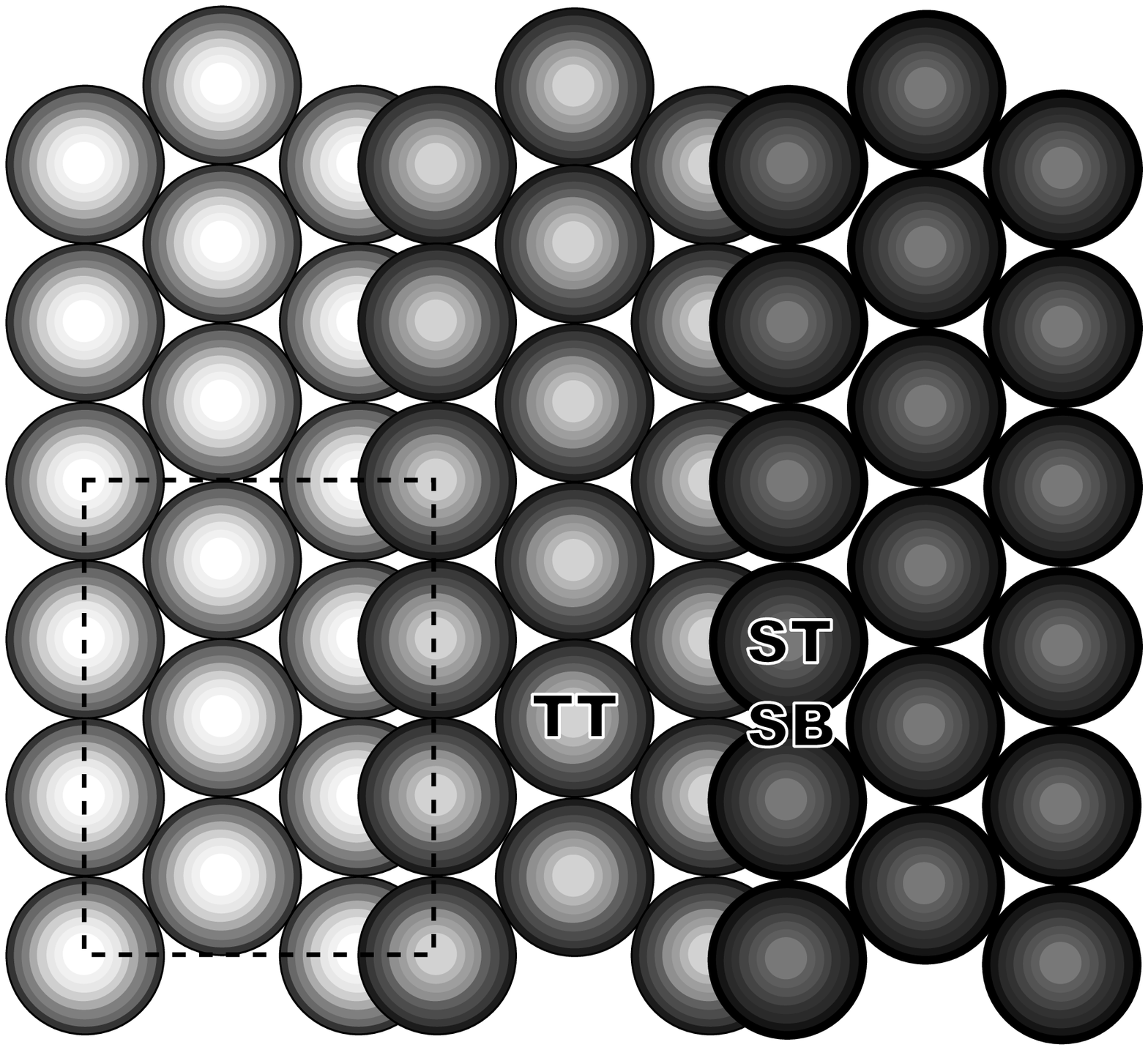}\hspace*{5mm}\includegraphics[width=4.0cm]{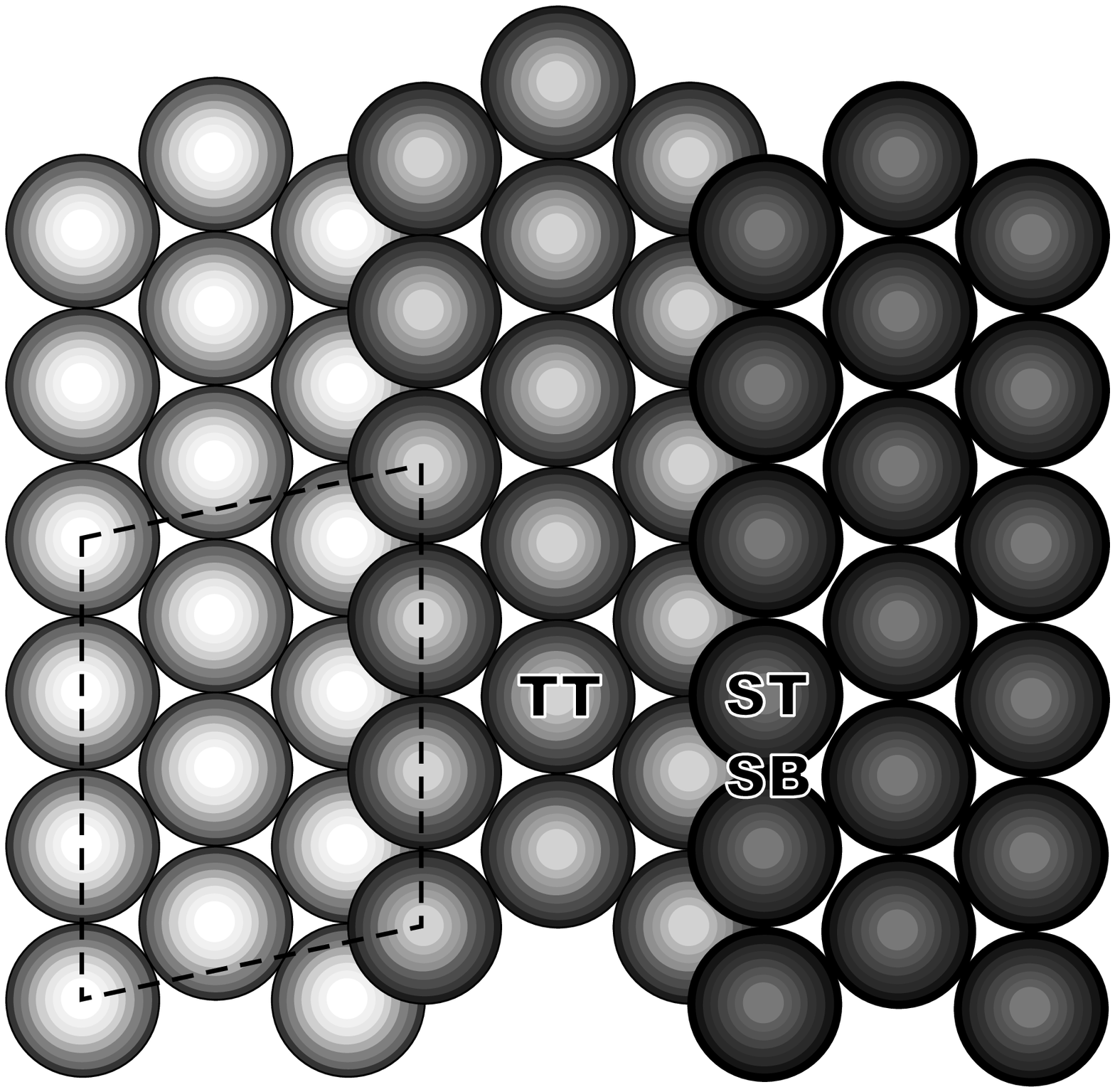}\\
\vspace*{-5mm}
\hspace*{1mm}(c) \hspace*{38mm} (d)\\
\vspace*{-10mm}
\includegraphics[width=3.5cm]{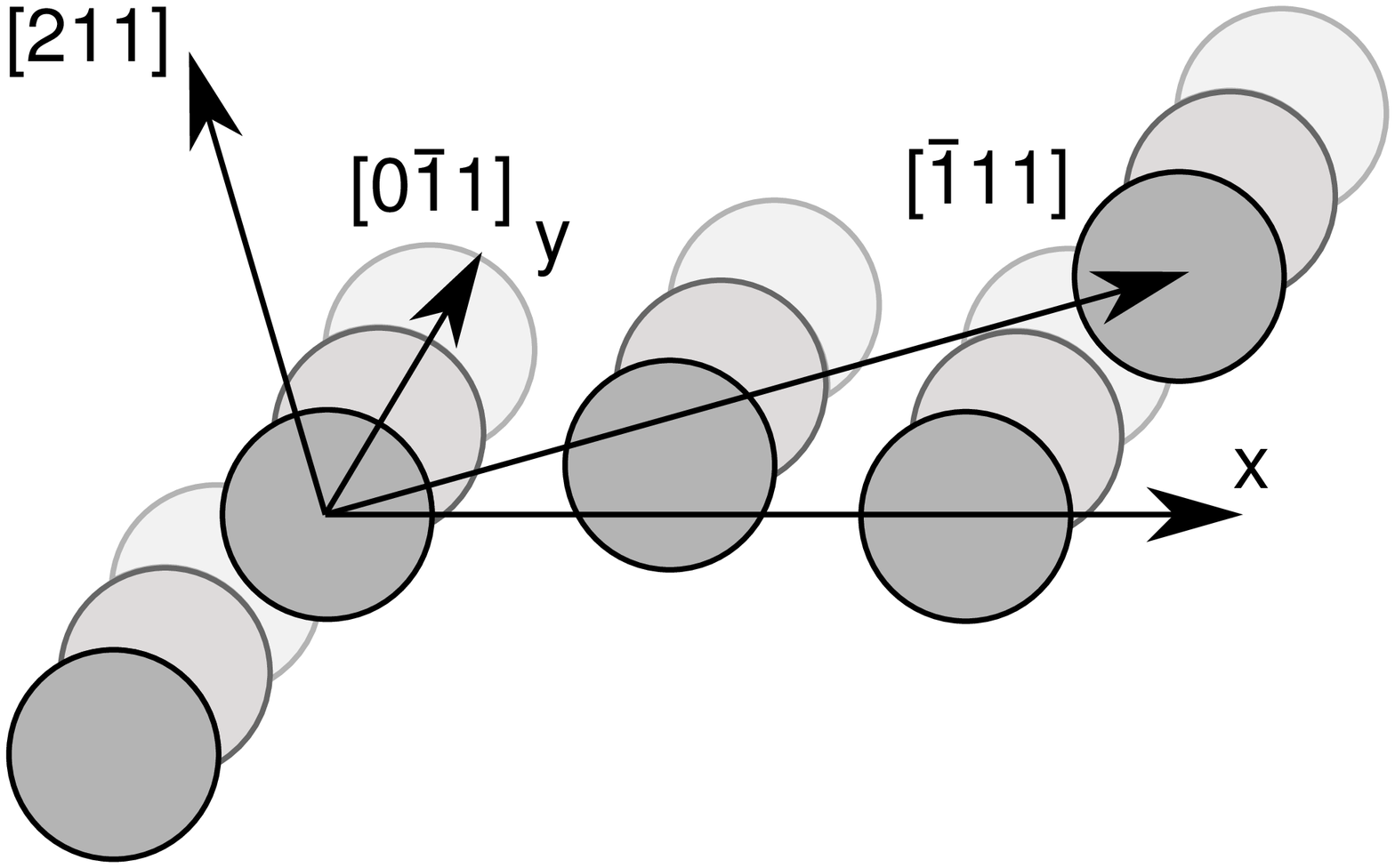}\hspace*{10mm}\includegraphics[width=3.5cm]{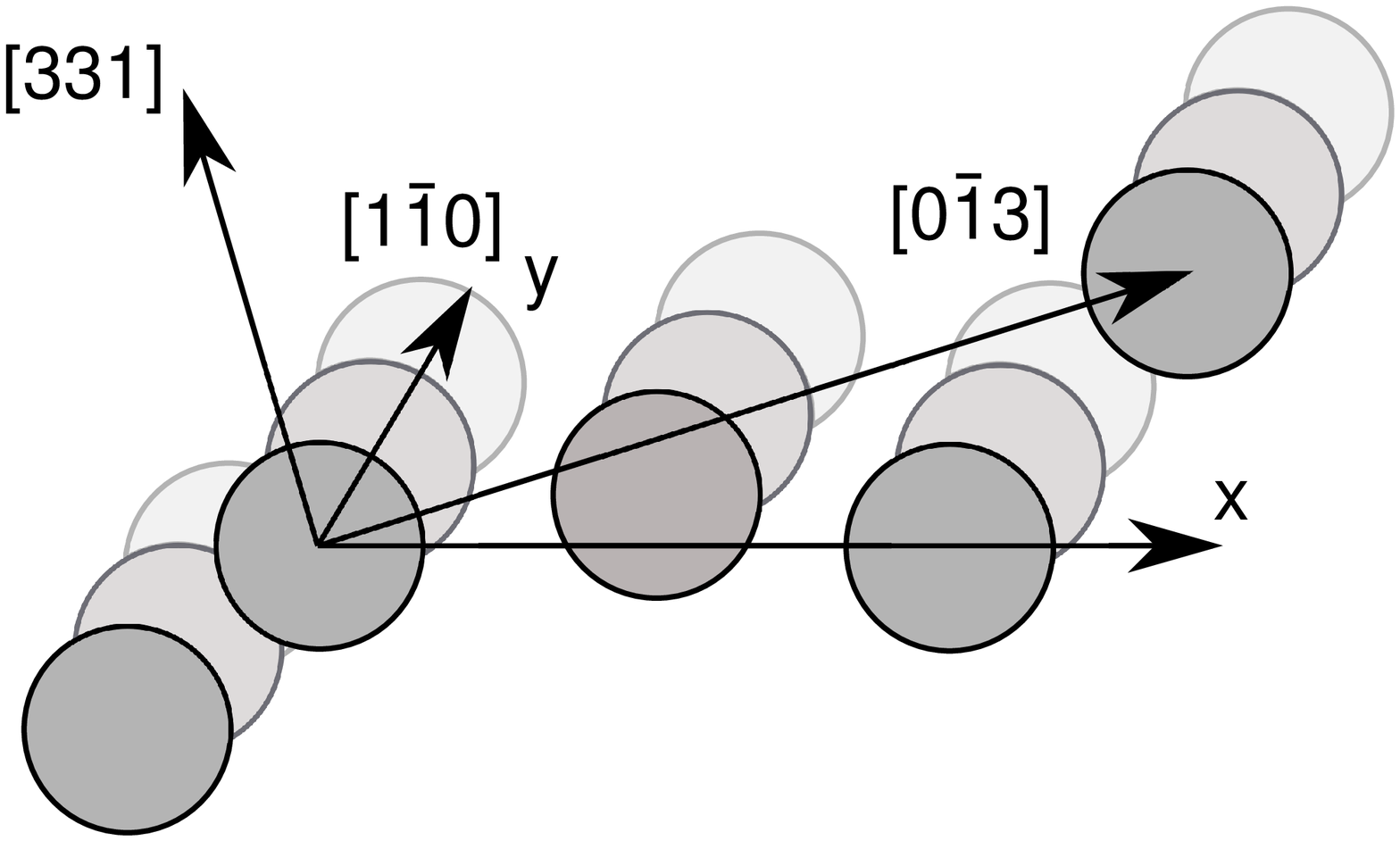}\\
\vspace*{-10mm}
\caption{\label{fig:structures}Top view of (a) Pt(211)
   and (b) Pt(331) surfaces. Terraces are both fcc(111), but
   the step edge of Pt(211) is fcc(100) ($A$ type) while the one
   for Pt(331) is fcc(111) ($B$ type). The surface supercells used
   in the calculations are shown with dashes lines. The most stabile
   adsorption sites for H are also shown: terrace-top (TT), step-top (ST),
   and step-bridge (SB) site.  Side view of (c) Pt(211) and (d) Pt(331)
   surfaces with corresponding lattice directions in the supercells. In addition, the
   $y$ axis is along the step edge and $x$ axis along the terrace
   and perpendicular to the step edge, {\it i.e.}\ $x$ points to $[\overline{2}11]$
   in (c) and to $[\overline{1}\overline{1}2]$ in (d).}
\end{figure}

Both the (211) and (331) vicinal surfaces consist of three-atoms
wide fcc(111) terraces. These vicinal surfaces have different
atomistic structures at the step edges. The step is called an {\it
A-type} step if the orientation of the step is (100) and a {\it
B-type} step if the orientation is (111). The ideal positions of the
atoms for the (211) and (331) surfaces are schematically shown in
Fig. \ref{fig:structures}.

\section{Results and discussion}

\subsection{Potential-Energy Surfaces}

We have calculated three-dimensional adiabatic potential-energy
surfaces (3D APES) for H on Pt(111), Pt(211) and Pt(331). The total
energy for H on Pt has been calculated in a grid on each surface and
at each site with different heights ($z$) around the minimum
position. The Morse potential \cite{MorXX} fitted to the data is of
the form
\begin{displaymath} 
U(x,y,z)=A(x,y)\left\{1-e^{-B(x,y)[z-C(x,y)]}\right\}^2+D(x,y),
\end{displaymath}
where $A$, $B$, $C$ and $D$ are fitting parameters that depend on
the position $(x,y)$. This potential form works well when the height
from the surface is over $0.3$ {\AA}. Under this limit the potential
is overestimated near the fcc and hcp sites. However, by studying
wave functions and eigenvalues of the H atom on Pt(111)
\cite{Bad03,Kal01,Nob01}, we conclude that it is adequate to
calculate the potential accurately using a dense grid near the
minimum at each $(x,y)$ site and elsewhere use a coarse grid and a
feasible approximation for the potential.

\subsubsection{H on the Flat Pt(111) Surface}

A full monolayer (1.0 ML) of H on Pt(111) has been calculated to
test that our present results are consistent with previous DFT
calculations. Also, the data for H on the flat Pt(111) surface is
needed to be able to see how a step changes the electronic structure
seen by the H atom. Even though the activation energy for H
diffusion depends on the H coverage \cite{Zhe04}, the 2D APES of H
on Pt(111) with 0.25 and 1.0 ML are qualitatively similar
\cite{Bad04}, thus we can use the present 1.0 ML data for H/Pt(111)
when compared the qualitative changes due to the step in the 2D APES
of H on stepped Pt(111) surfaces.

\begin{figure}
(a)\\
\vspace*{-20mm}\includegraphics[width=8.0cm]{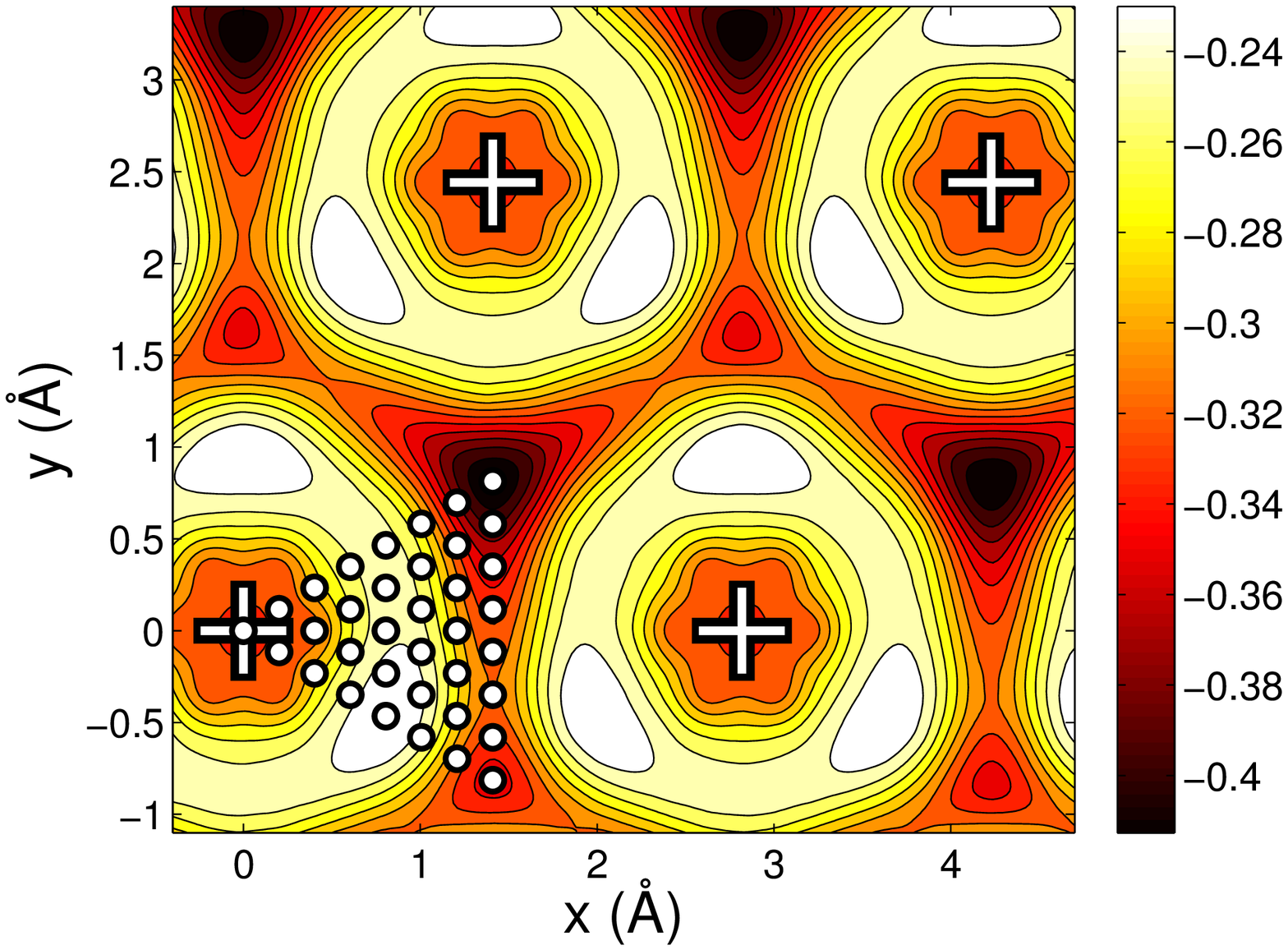}\\
\vspace*{-25mm}(b)\\
\vspace*{-25mm}\includegraphics[width=8.0cm]{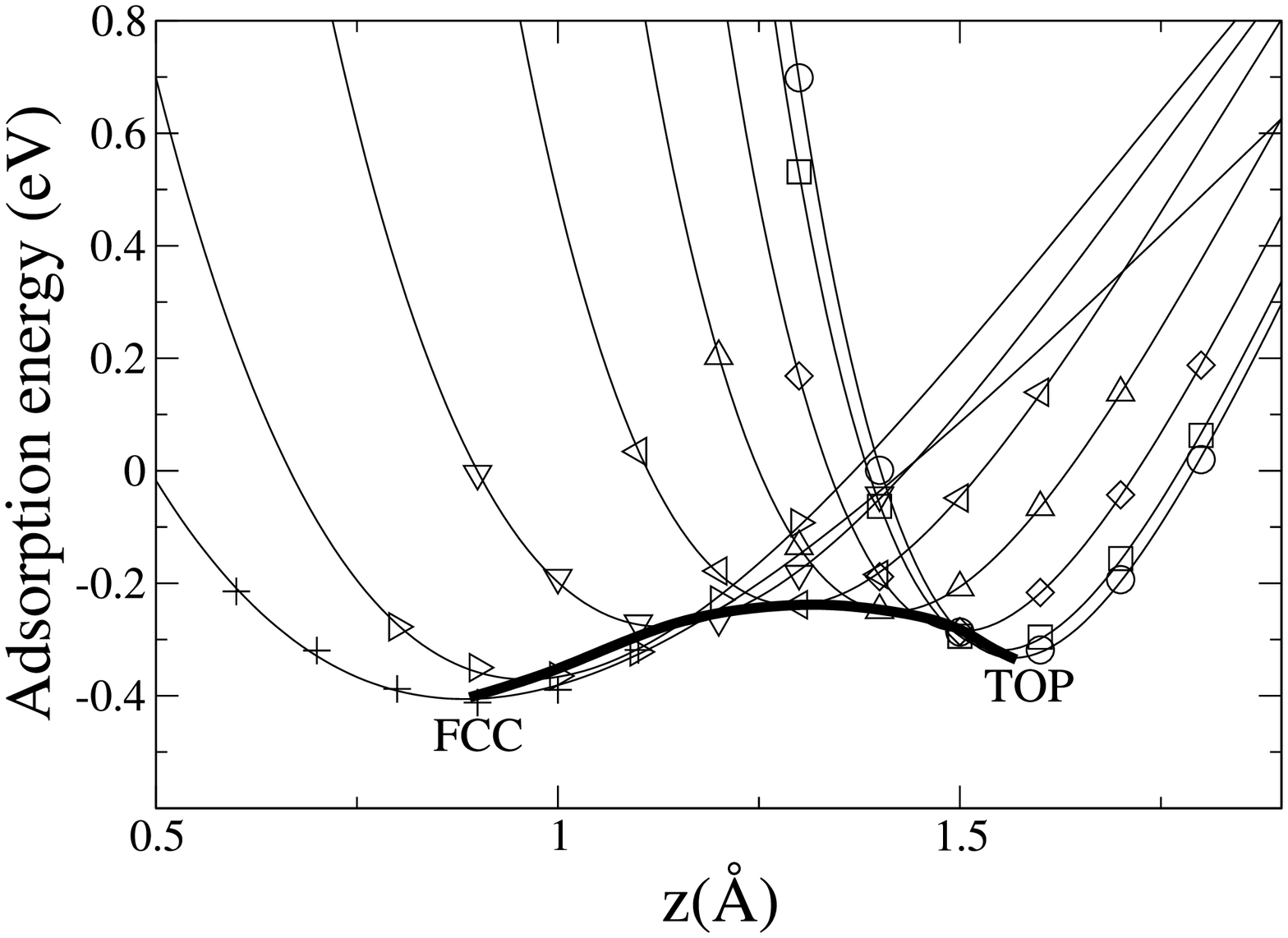}\\
\vspace*{-25mm}(c)\\
\vspace*{-25mm}\includegraphics[width=8.0cm]{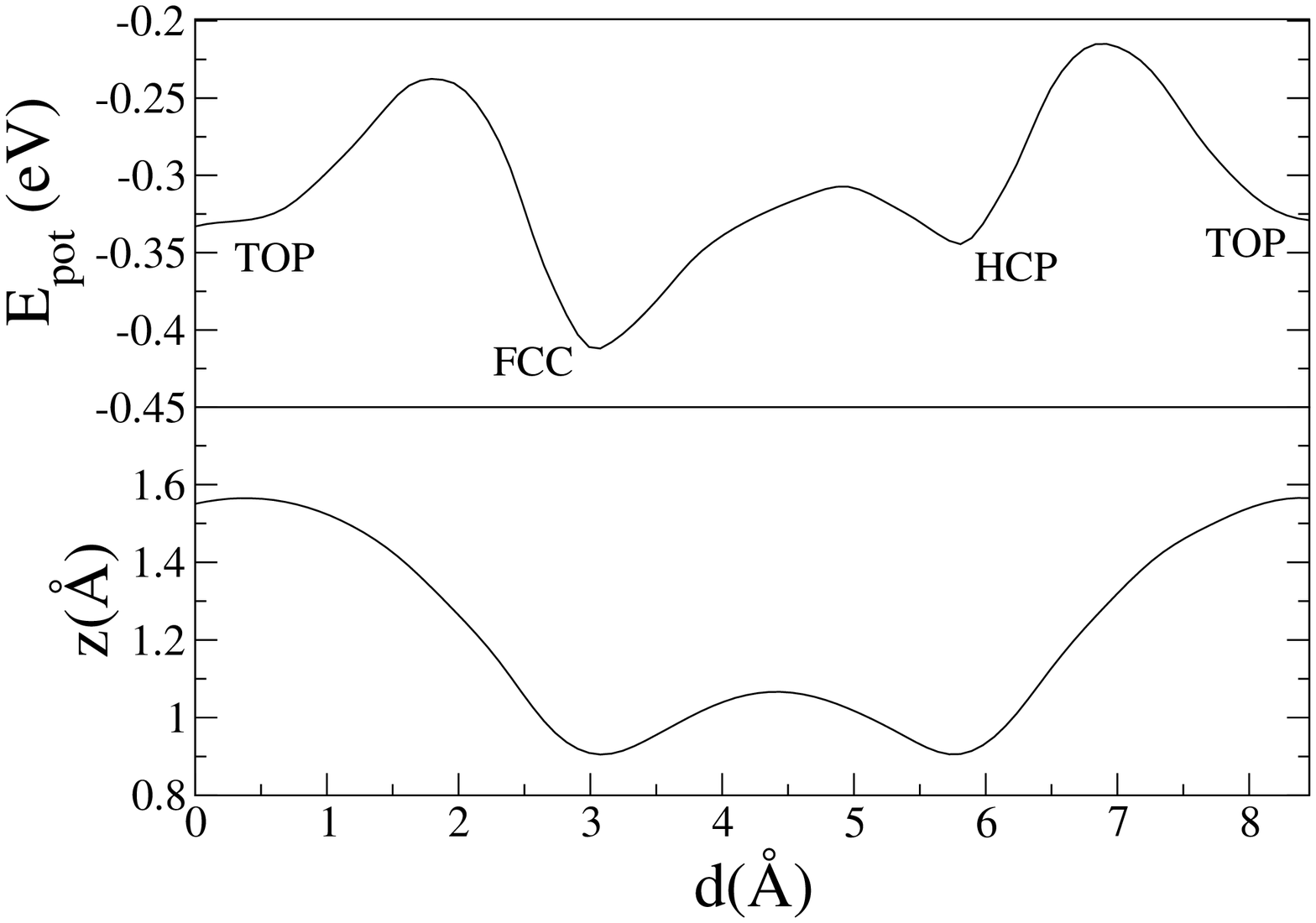}\\
\vspace*{-25mm}
 \caption{\label{fig:apes-pt111}
   Results for H on Pt(111). (a) (Color online) Minimum energy 2D APES for H on Pt(111) (in eV).
   Crosses represent the positions of surface Pt atoms and open dots the grid points
   used in the calculations. The $x$ and $y$ axis corresponds to
   $[\overline{1}10]$ and $[\overline{1}\overline{1}2]$ directions,
   respectively.
   (b) Energy for different $z$ positions around the minimum
   along the path from an fcc to top site.
   (c) Minimum energy and height of H along the path top--fcc--hcp--top
   as a function of the distance $d$ along the path.}
\end{figure}

In Fig.\ \ref{fig:apes-pt111}(a) we show the calculation grid with a
total number of 36 equidistant positions on the $(x,y)$ plane over
the surface. For each of these positions the total energy of the
system has been calculated using various $z$ values around the
minimum position, the total number of calculated points being 219.
The minimum energy 2D APES is shown in Fig.\
\ref{fig:apes-pt111}(a), the total energy of the system with H at
different heights, \emph{i.e.}\ $z$ positions around the minimum
along the path from an fcc to top site in Fig.\
\ref{fig:apes-pt111}(b), and the minimum energy and height of H
along the path top--fcc--hcp--top are presented in Fig.\
\ref{fig:apes-pt111}(c).

In Table \ref{table:pt111}, the adsorption energy of the H atom at
different sites on the Pt(111) surface, the energy differences for H
to move from a site to another (see also Fig.\
\ref{fig:apes-pt111}(c)), and calculated vibrational energies based
on the harmonic approximation for perpendicular ($\hbar
\omega_{\perp}$) and parallel motion ($\hbar \omega_{\parallel}$)
are presented.
\begin{table*}
 \caption{\label{table:pt111}
      H on Pt(111). Adsorption energy $E_{ads}$, activation energy $\Delta E$
      (in parenthesis the site towards which the barrier is represented),
      harmonic vibrational energies $E_{vib}^\perp$ and $E_{vib}^\parallel$
      along directions $[hkl]$. All data in eV.}
 \begin{ruledtabular}
  \begin{tabular}{lcccccc}
   site & $E_{ads}$ & \multicolumn{2}{l}{$\Delta E$ (neighboring site)}  & $E_{vib}^\perp [111] $ & $E_{vib}^\parallel [\overline{1}10] $ & $E_{vib}^\parallel [\overline{1}\overline{1}2] $\\\hline
   top   & $-$0.33  &  0.090 (fcc) & 0.120 (hcp) & 0.277 & 0.066 & 0.033  \\
   hcp   & $-$0.35  &  0.035 (fcc) & 0.140 (top) & 0.144 & 0.099 & 0.028   \\
   fcc   & $-$0.42  &  0.200 (top) & 0.103 (hcp) & 0.143 & 0.082 & 0.082   \\
  \end{tabular}
 \end{ruledtabular}
\end{table*}
The calculated energies represented in Table \ref{table:pt111} and
the energy profile of the 2D APES along the path top--fcc--hcp--top
for H on Pt(111) shown in Fig.\ \ref{fig:apes-pt111}(c) can be
compared to those calculated by B\u adescu {\it et al}.\
\cite{Bad03} and Olsen {\it et al}.\ \cite{Ols99}. The curvature and
shape of the energy profile are very similar compared to those of
B\u adescu {\it et al}. The only major difference is the top site
where the energy is much higher in the present study. However, it is
highly unlikely that H is adsorbed at the top site \cite{Bad03} and
thus one cannot resolve further the energy of H at the top site
based on the experiments. To check the accuracy numerically we made
some test calculations with a different GGA (PBE \cite{Per96}),
potential (ultrasoft pseudopotential \cite{Van90}) and Fermi
smearing method. The difference between the energies of H at the fcc
and hcp sites is hardly affected at all, the difference in all the
cases being 0.07 eV, while there are some variations in the energy
difference between fcc and top. With PBE GGA the energy difference
between H on Pt(111) at the fcc and top sites is 0.10 eV, with
ultrasoft pseudopotetial 0.11 eV, and with Fermi smearing it is 0.08
eV, the energy difference with PW91 GGA being 0.09 eV (Table
\ref{table:pt111}). Thus, the error due to different methods is
about 0.02 eV and the present results seem to be quite accurate.

There are more differences in the energy profile when compared to
that of Olsen {\it et al}.\ \cite{Ols99}. Nevertheless, the
vibrational energies based on the harmonic approximation are in a
good agreement both with the results by Olsen {\it et al}.\
\cite{Ols99} and B\u adescu {\it et al}.\ \cite{Bad03}. This
indicates similar curvature and shape of the 2D potential-energy
surfaces for H at the high-symmetry sites on Pt(111).

\subsubsection{H on the Vicinal Pt(211) and Pt(331) Surfaces}

\begin{figure}
(a)\\
\vspace*{-20mm}\includegraphics[width=8.0cm]{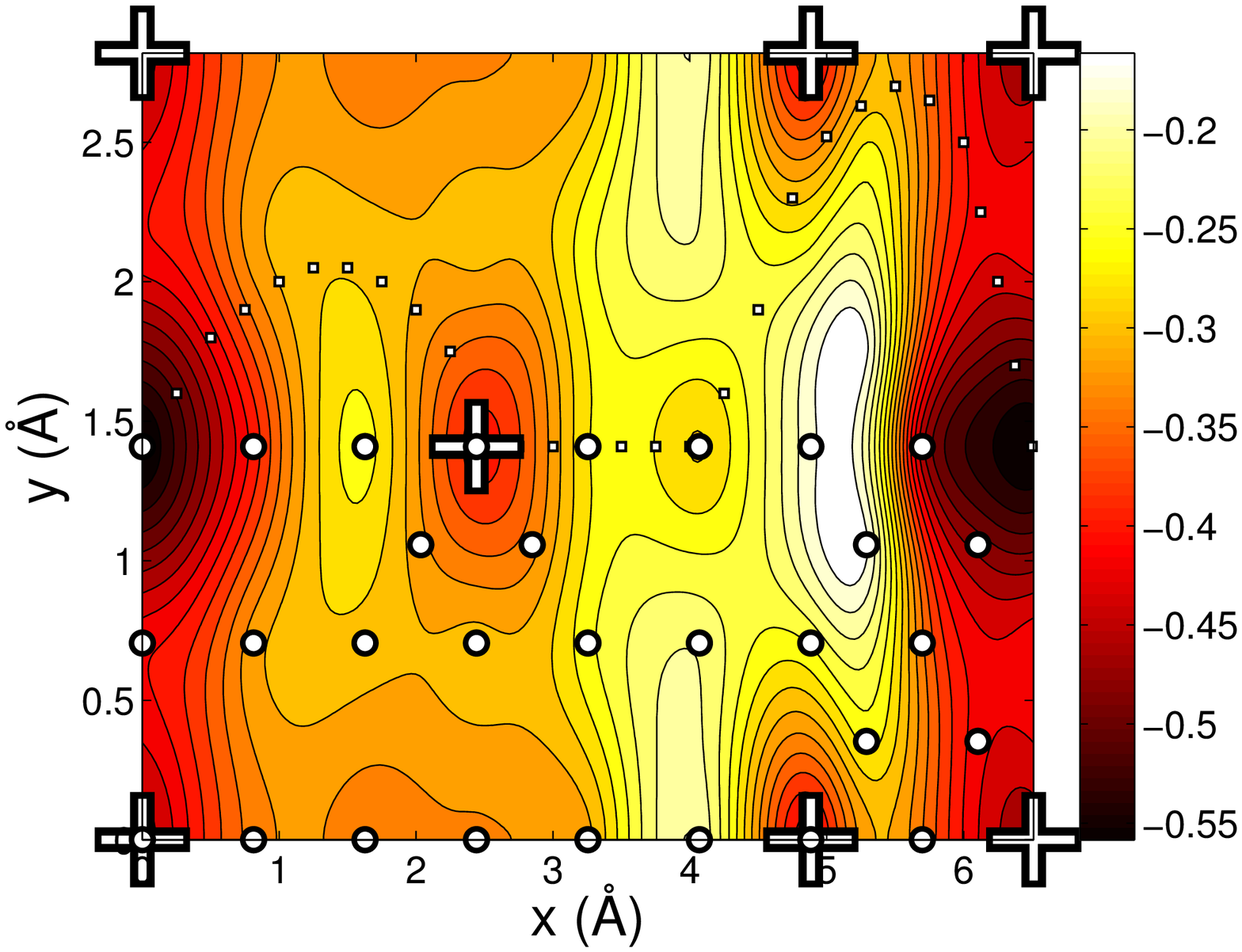}\\
\vspace*{-25mm}(b)\\
\vspace*{-25mm}\includegraphics[width=8.0cm]{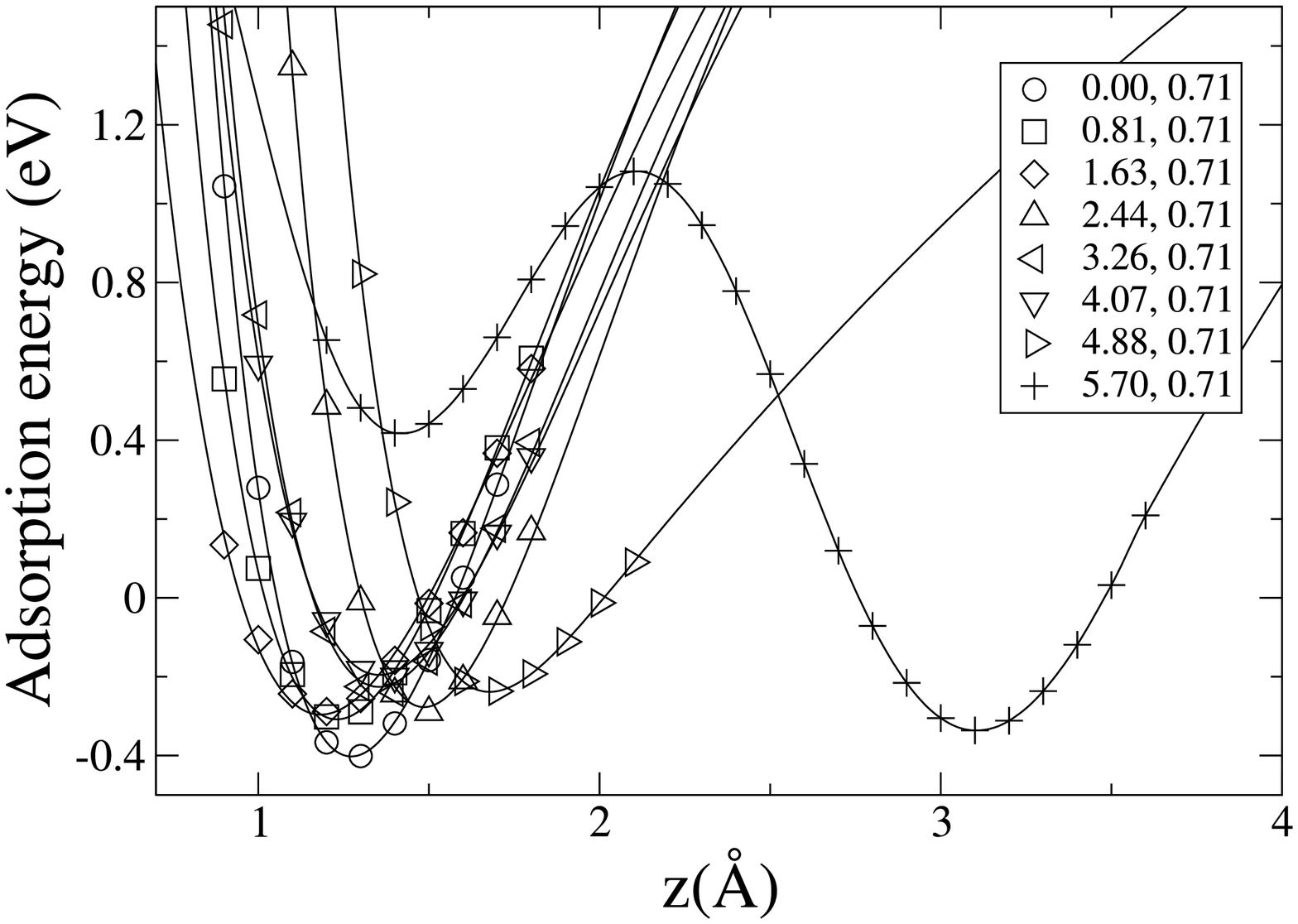}\\
\vspace*{-25mm}(c)\\
\vspace*{-25mm}\includegraphics[width=8.0cm]{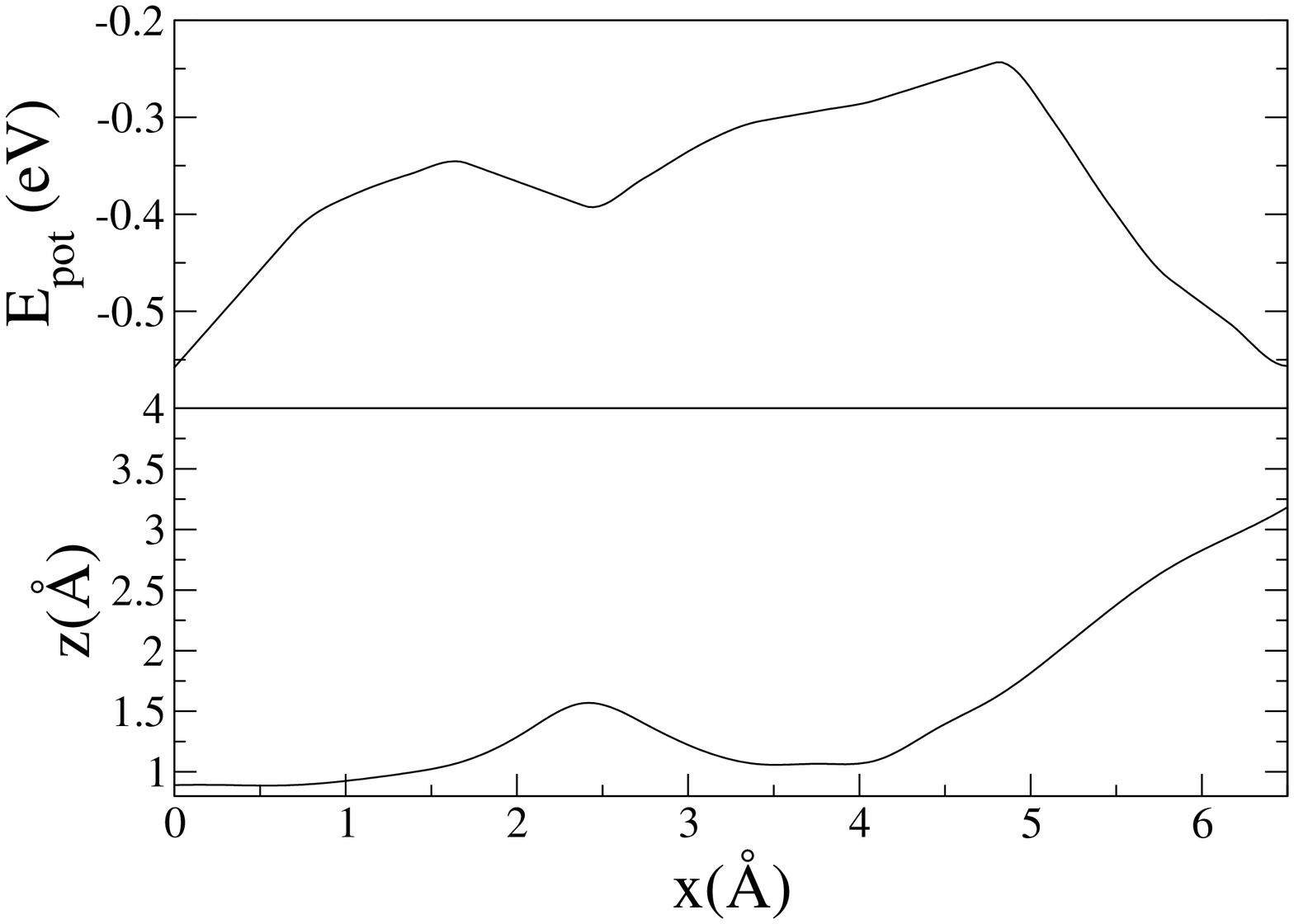}\\
\vspace*{-25mm}
 \caption{\label{fig:apes-pt211}
   Results for H on Pt(211). (a) (Color online) Minimum energy 2D APES for H on the surface (in eV).
   Crosses represent the positions of surface Pt atoms such that the two right-most atoms are
   the edge atoms on the upper terrace, the other ones residing on the lower terrace.
   Open dots are the grid points used in the calculations. Small open squares represent the minimum
   energy path across the surface perpendicular to the step edge.
   (b) Energy for different $z$ positions around the minimum
   along the path whose coordinates ({\AA}) are included in the insert.
   (c) Minimum energy profile and height of H along the minimum energy path shown in (a).}
\end{figure}

\begin{figure}
(a)\\
\vspace*{-20mm}\includegraphics[width=8.0cm]{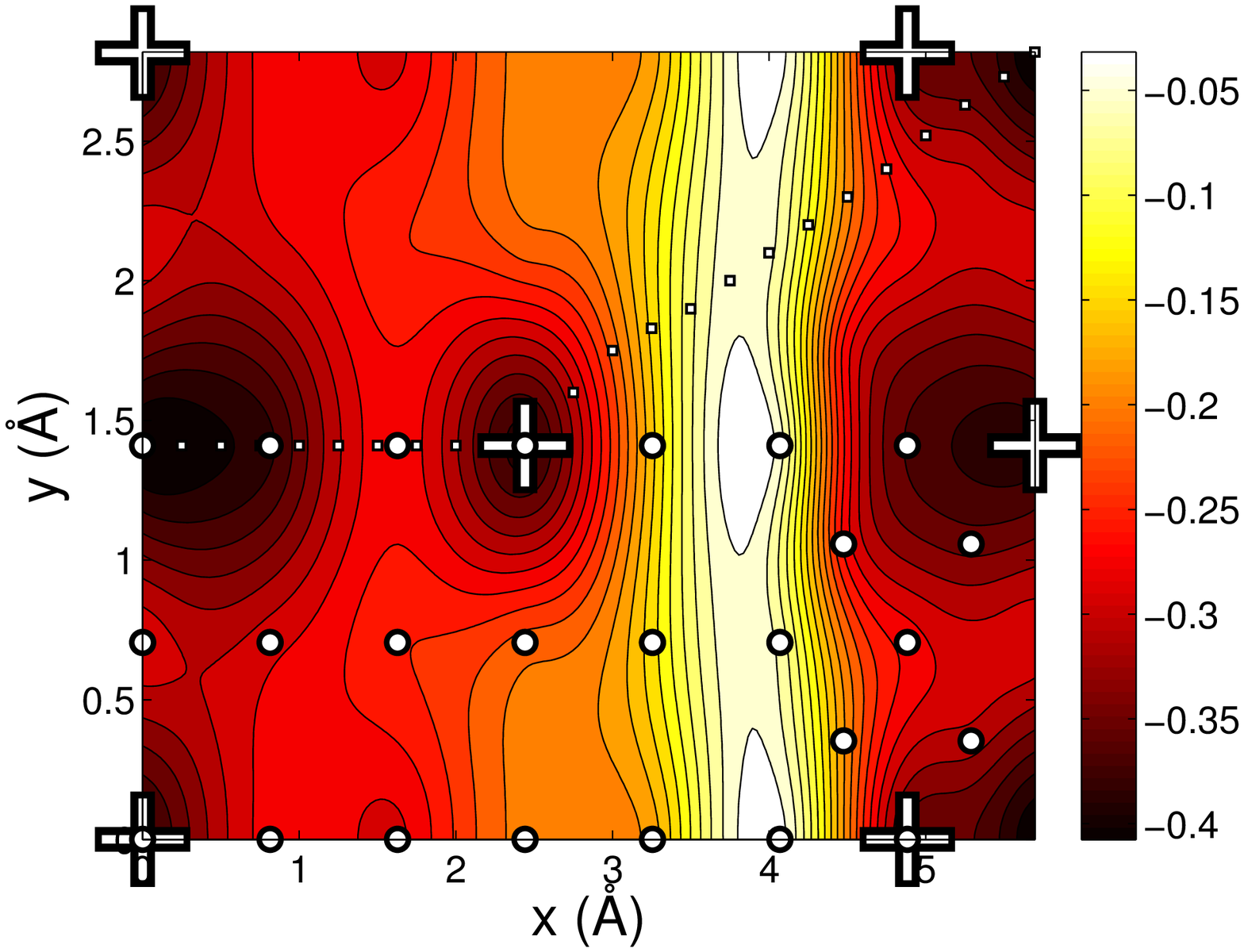}\\
\vspace*{-25mm}(b)\\
\vspace*{-25mm}\includegraphics[width=8.0cm]{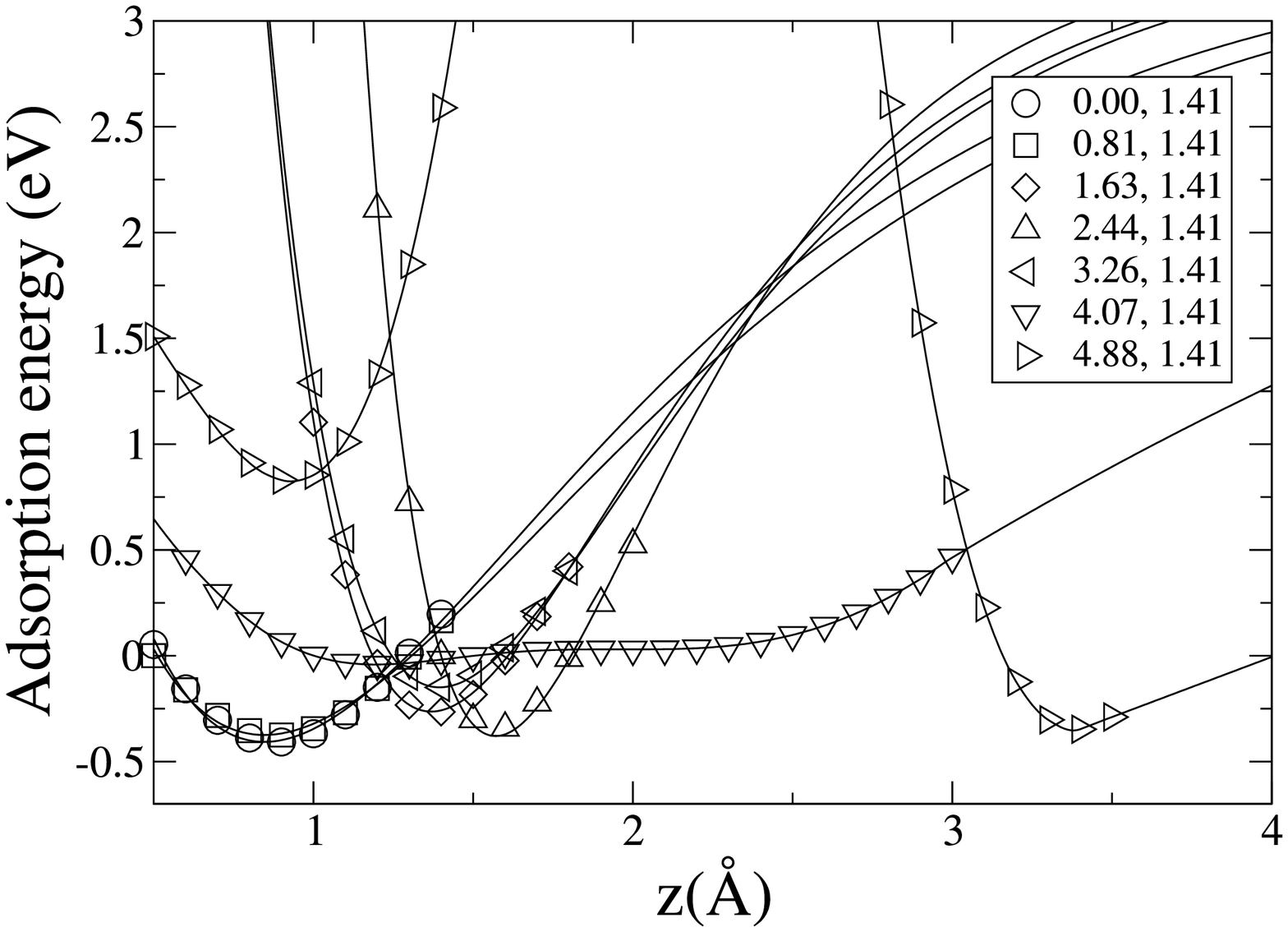}\\
\vspace*{-25mm}(c)\\
\vspace*{-25mm}\includegraphics[width=8.0cm]{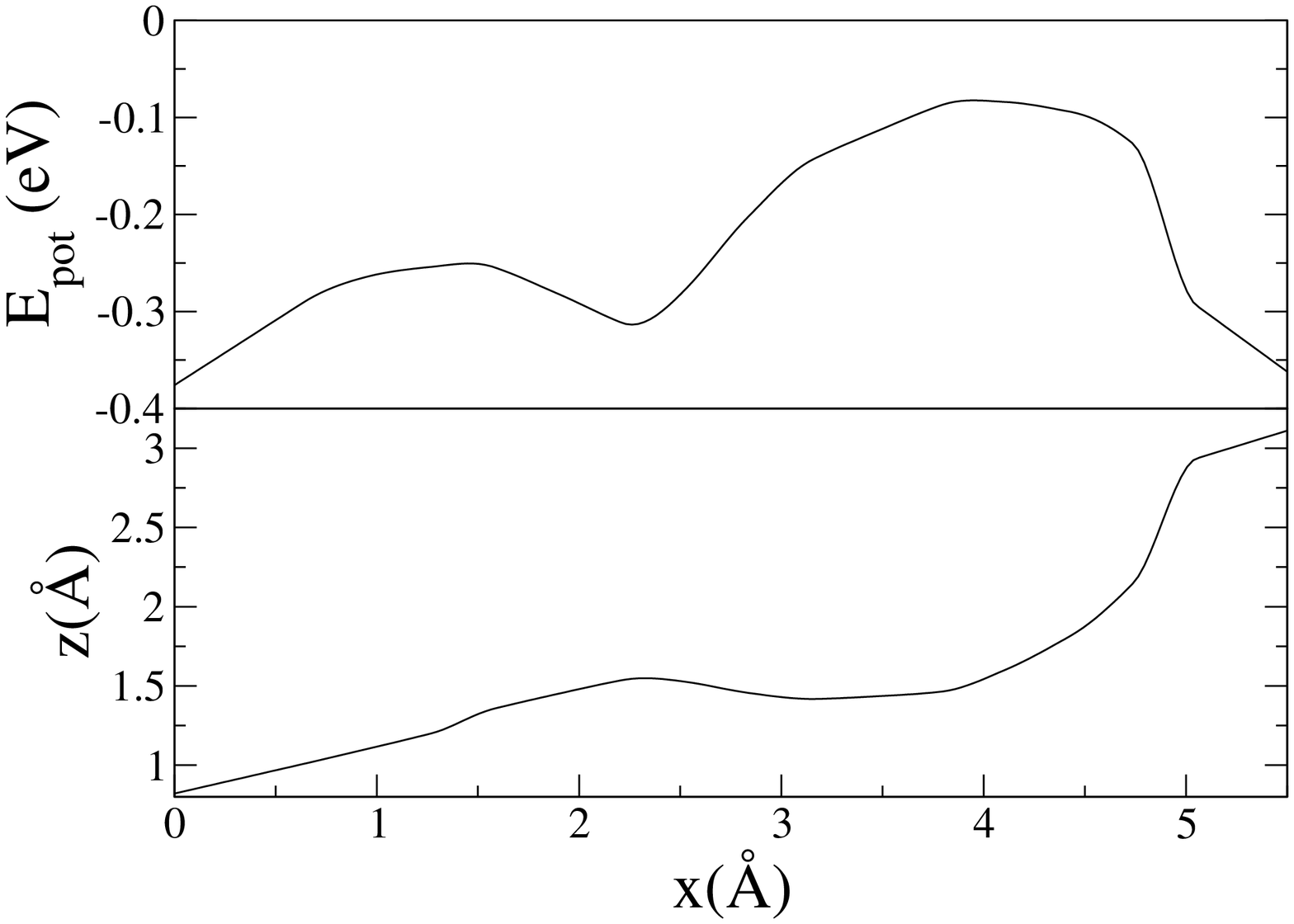}\\
\vspace*{-25mm}
 \caption{\label{fig:apes-pt331}
   Results for H on Pt(331). See details in the caption of Fig.\ \ref{fig:apes-pt211}.
   {\it N.B.} In (a) the Pt atom at $(0,0)$ corresponds to the one on the upper terrace
   at $(5.69,1.41)$. See also Fig.\ \ref{fig:structures}(b).}
\end{figure}

The 3D APES's of H on Pt(211) and Pt(331) have been calculated using
a grid consisting of 30 positions for H on Pt(211) and 25 positions
for H on Pt(331) on the $(x,y)$ plane over the surface. The grids
are constructed so that there is a grid point at every high-symmetry
site (fcc, hcp, top and bridge) and at least one midpoint between
these sites. This assures that the adsorption energy at each local
minimum site is calculated and also the energy barrier between these
sites is calculated with a reasonable accuracy. The grids used in
the calculations are shown in Figs.\ \ref{fig:apes-pt211}(a) and
\ref{fig:apes-pt331}(a). The total energy of the system has been
calculated at each surface site using various $z$ values. The number
of calculated $z$ values is $7-10$ at terrace sites and $20-30$ at
sites near the step edge. This difference is due to the shape of the
potential. At terrace sites the geometry of the potential is similar
to that of the Morse potential and the number of calculated points
can be small, but near the step edge the shape is more complicated
due the interaction between the H atom and step atoms, and thus a
denser grid along the $z$ direction near the step region is needed.

The calculated 2D APES's of H on Pt(211) and Pt(331) are also shown
in Figs.\ \ref{fig:apes-pt211}(a) and \ref{fig:apes-pt331}(a),
respectively. There are three energy minima for H on the Pt(211) and
Pt(331) surfaces. The deepest minimum is between two step-edge atoms
({\it step-bridge site}) for both the Pt(211) and Pt(331) surfaces.
The other two minima are above a terrace atom ({\it terrace-top
site}) and above a step-edge atom ({\it step-top site}).

The values of the adsorption and harmonic vibrational energies for H
on Pt(211) and Pt (331) are shown in Tables \ref{table:pt211} and
\ref{table:pt331}, respectively. There are only minor changes for
the perpendicular vibrational energies; it is mainly the parallel
vibrational models that are affected by the steps. On the other
hand, when we compare these 2D APES plots to the 2D APES plot for H
on the ideal Pt(111) surface, one cannot see any similarities in the
(111) terrace region. The step thus influences the adsorption
everywhere in the three-atom wide terrace region.

\begin{table}
 \caption{\label{table:pt211}H on Pt(211). Adsorption energy $E_{ads}$, harmonic vibrational energies
        $E_{vib}^\perp$ and $E_{vib}^\parallel$ along directions $[hkl]$. All data in eV.}
 \begin{ruledtabular}
 \begin{tabular}{lcccc}
  site & $E_{ads}$ & $E_{vib}^\perp [211] $ & $E_{vib}^\parallel [\overline{1}11] $ & $E_{vib}^\parallel [0\overline{1}1] $ \\\hline
   step-top    & $-$0.44  &  0.269 & 0.048  & 0.046  \\
   step-bridge & $-$0.59  &  0.160 & 0.069  & 0.126  \\
   terrace-top & $-$0.38  &  0.268 & 0.088  & 0.048  \\
 \end{tabular}
 \end{ruledtabular}
\end{table}

\begin{table}
 \caption{\label{table:pt331}H on Pt(331). Adsorption energy
        $E_{ads}$, harmonic vibrational energies
        $E_{vib}^\perp$ and $E_{vib}^\parallel$ along directions $[hkl]$. All data in eV.
        Direction $[\overline{1}\overline{1}6]$ is along the $(331)$ surface and perpendicular
        to the step-edge direction $[1\overline{1}0]$. See also Fig.\ \ref{fig:structures}.}
 \begin{ruledtabular}
  \begin{tabular}{lcccc}
  site & $E_{ads}$ & $E_{vib}^\perp [331] $ & $E_{vib}^\parallel [\overline{1}\overline{1}6] $ & $E_{vib}^\parallel [1\overline{1}0] $ \\\hline
   step-top    & $-$0.37  &  0.268 & 0.039  & 0.051  \\
   step-bridge & $-$0.41  &  0.162 & 0.046  & 0.120  \\
   terrace-top & $-$0.36  &  0.265 & 0.096  & 0.061  \\
  \end{tabular}
 \end{ruledtabular}
\end{table}

The calculated 2D APES for H on Pt(211) can be compared to the one
calculated by Olsen {\it et al}.\ \cite{Ols04}. These two are in
close agreement and the differences can be explained with different
computational parameters and interpolation methods. Also, the
calculation grid used in the present work is denser than the grid
used by Olsen {\it et al}.

The effect of the step edge can also be seen in the behavior of the
total energy as a function of the height of H from the surface near
the step edge. The calculated total energy of H on stepped Pt
surfaces at selected positions (for more data see the web link in
Ref.\ \cite{webpage}) as a function of the height $z$ from the
surface is plotted in Figs.\ \ref{fig:apes-pt211}(b) and
\ref{fig:apes-pt331}(b). Near the step edge the potential has a
double-well structure not seen in the 2D APES's, but which can be
seen when the height of the H atom is modified. When the H atom is
far away from the step edge the shape of the potential in the $z$
direction is of the Morse potential form. The double well structure
of the potential exists near the step region and it is due the
interaction between the step atoms and the H atom. It is also clear
that there are significant quantitative differences between the full
3D APES of the (211) and (331) surfaces. For example, the $z$
positions and the curvatures of the minima differ dramatically. Such
differences in the APES lead into large differences in the
vibrational band structure of H adatoms on these surfaces, which
indicates significantly different behavior for quantum diffusion of
H at low temperatures.

The minimum-energy paths (MEP) for H moving from a lower terrace to
an upper one were calculated using the nudged elastic band (NEB)
method \cite{Jon98} with which one can find (local) minimum path for
transition between two local minimum configurations. The
corresponding classical diffusion paths for H on Pt(211) and Pt(331)
are presented in Figs.\ \ref{fig:apes-pt211}(a) and
\ref{fig:apes-pt331}(a), respectively. The corresponding energy
values and the distance of the H from the surface are presented in
Figs.\ \ref{fig:apes-pt211}(c) and \ref{fig:apes-pt331}(c),
respectively. In these plots one can see a potential barrier when H
approaches the step. This barrier is located before the step edge of
the upper terrace and before the formation of the double-well
potential. The energy barriers between two step-bridge minima are
0.30 and 0.29 eV for the activated diffusion of H on the Pt(211) and
Pt(331) surfaces, respectively. On the other hand, the energy
barriers along the step edges are lower, being only 0.15 and 0.11 eV
for H on the Pt(211) and Pt(331) surfaces, respectively.

\subsection{Local Density of States}

To understand better the bonding between the H atom and Pt surface
atoms we have calculated the local density of states (LDOS) for H at
selected minimum energy sites and for the nearest Pt atoms and
projected them onto the states of the chosen angular momentum. The
Wigner-Seitz radii used for H and Pt atoms are 0.370 {\AA} and 1.455
{\AA}, respectively. An extensive set of calculated LDOS's can be
found on the web page in Ref. \cite{webpage}.

\begin{figure}
(a)\\
\vspace*{-20mm}\includegraphics[width=8.0cm]{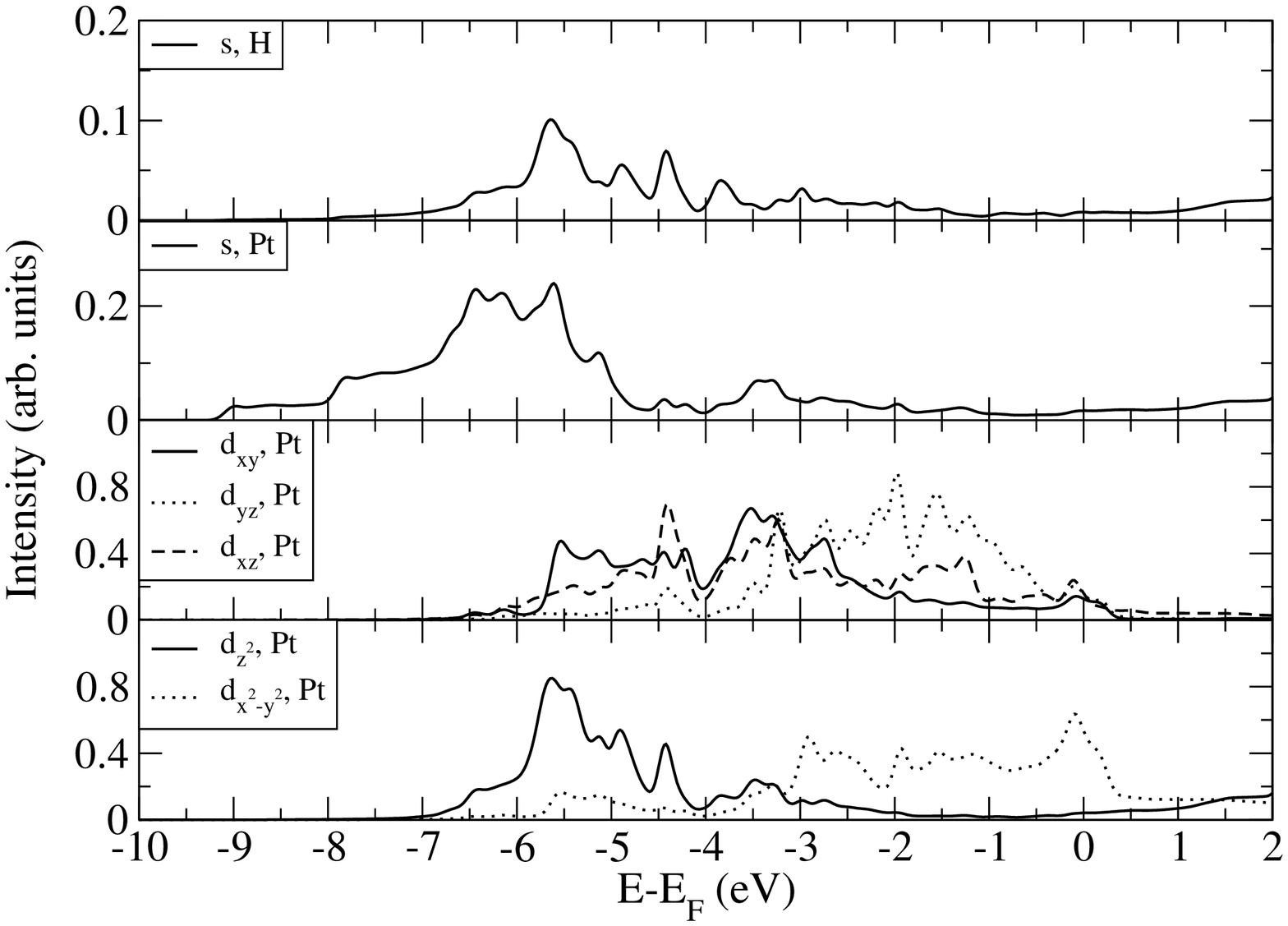}\\
\vspace*{-25mm}(b)\\
\vspace*{-25mm}\includegraphics[width=8.0cm]{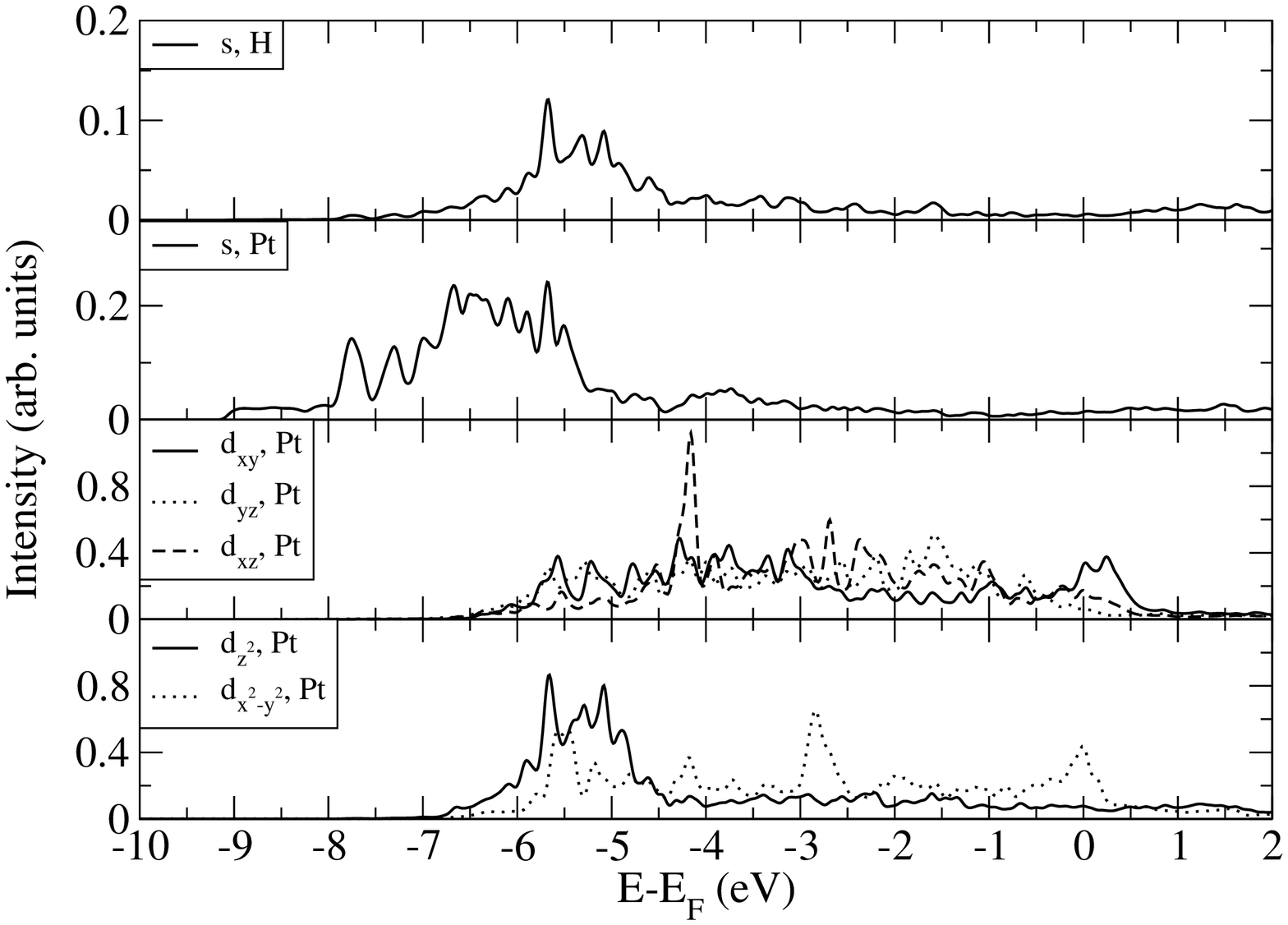}\\
\vspace*{-25mm}(c)\\
\vspace*{-25mm}\includegraphics[width=8.0cm]{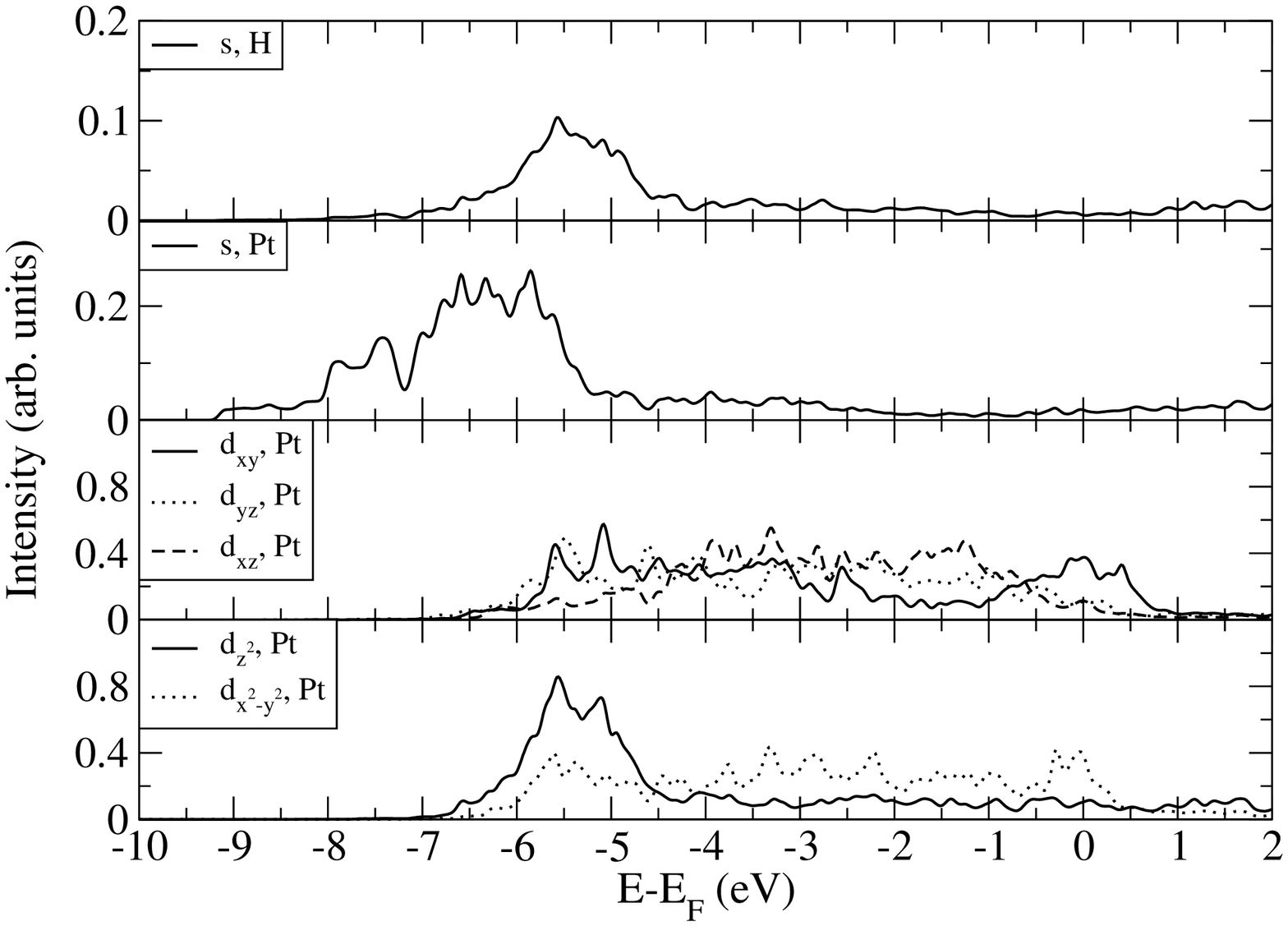}\\
\vspace*{-25mm}
 \caption{\label{fig:dos-pt}
  LDOS curves for H on Pt surfaces: H at (a) the top site on Pt(111),
  (b) the terrace-top site on Pt(211), and (c) the terrace-top site
  on Pt(331).}
\end{figure}

There are three minimum energy sites for H on the Pt(111) surface,
\emph{i.e.} fcc, hcp and top. LDOS's have been calculated at each of
these sites for H and the nearest Pt atom of H \cite{webpage}. For H
adsorbed at the top site the 1s orbital of H and 5d$_{z^2}$ of Pt
hybridize in the energy range of [$-7$ eV, $-3$ eV] (see Fig.\
\ref{fig:dos-pt}(a)). The adsorption of H at the fcc and hcp hollow
sites is similar, but considerably different from the adsorption of
H at the top site \cite{webpage}. For H at the fcc and hcp sites,
new peaks appear in LDOS's between [$-8$ eV, $-7$ eV]. These LDOS's
show hybridization of the 1s orbital of H and the 6s and 5d$_{xy}$
orbitals of Pt. Hong {\it et al}.\ \cite{Hon05} and L\'egar\'e
\cite{Leg04} have calculated LDOS for H interacting with Pt(111)
surface and our results are in a good agreement with their results.

On the vicinal surfaces, there are three energy minima for H on
Pt(211) and Pt(331). The deepest minimum is at the step-bridge site
while the other two minima are at the terrace-top and step-top
sites. In Fig.\ \ref{fig:dos-pt} we show a series of LDOS's for the
flat (111) and vicinal surfaces for the top and terrace-top site,
respectively.
It can be clearly seen in Fig.\ \ref{fig:dos-pt}
that the steps influence the H and Pt $s$ orbitals relatively
little. However, there are significant differences between the more
localized $d$ orbitals of the Pt surface atoms. This indicates the
non-local nature of the influence of the steps even on the terrace
atoms on vicinal surfaces. Further analysis of the LDOS curves
indicates that for terrace-top and step-top sites there is
hybridization between the 1s orbital of H and the 5d$_{z^2}$ orbital
of Pt as on the Pt(111) surface when H is adsorbed at the top site.
When H is adsorbed at the step-bridge site the bonding and
electronic structure are different. LDOS shows strong hybridization
between the 1s orbital of H and the 6s, 5d$_{xz}$ and 5d$_{x^2-y^2}$
orbitals of the step-edge Pt atom. Thus, despite the geometric
differences between the $A$ and $B$ type of steps, the binding of H
to the Pt atoms on the surfaces seems to be similar.

\section{Summary and Conclusions}

In this work, we have used the first-principles DFT approach to
study the electronic structure of H adsorbed on the vicinal Pt(211)
and Pt(331) surfaces. We have evaluated the full 3D potential-energy
surface for the adsorbed hydrogen. We have also studied the local
density of states for the adsorbed hydrogen at various sites which
provides a more detailed look at the changes in electronic structure
due to the adsorption of hydrogen at various substrate sites. We
have found that the preferred adsorption (minimum energy) site for H
on Pt surfaces is the fcc site on the Pt(111) surface and the
step-bridge site, \emph{i.e.}\ the site between two step atoms, on
the Pt(211) and Pt(331) surfaces. The step edge on Pt surfaces
changes the potential-energy surface for H on the three-atoms wide
(111) terrace quite dramatically when compared to the case of the
flat Pt(111) surface. This is in qualitative agreement with the
experimental observation that the step effect on the electronic
structure is nonlocal and extends far beyond to the terrace location
\cite{Zhe04}. For instance, near the step edge the potential has a
double-well structure which is due to the interaction between the H
atom and Pt atoms of the step edge. Even though the adsorption
energies and the density of states as well as the vibrational
energies based on the harmonic approximation are quite similar for
the H on the Pt(211) and Pt(331) surfaces at the corresponding
adsorption sites, the anharmonic effects, for instance, can
contribute quite dramatically to the vibrational band structure and
also to the nature of diffusion of H on stepped surfaces. The
calculated 3D APES provides a necessary input for further studies of
vibrational and diffusive dynamics of H adatoms on the vicinal
surfaces of Pt.

\section*{Acknowledgements} This work has been in part supported by the
Academy of Finland through its Center of Excellence program (COMP).
We greatly acknowledge generous computing resources provided by the
Finnish IT Center for Science (CSC).

\end{document}